\documentclass{aa}
\usepackage{graphicx}
\begin{document}
   \title{A young and complex binary star - HD 144432}

   \subtitle{}

   \author{M. R. P\'erez 
          \inst{1}
          \and
          M. E. van den Ancker\inst{2}\fnmsep\thanks{Now at European Southern Observatory, 
          Karl-Schwarzschild-Strasse 2, D-85748, Garching bei M\"unchen, Germany}
          \thanks{Visiting Astronomer at the Infrared Telescope Facility, 
          which is operated by the University of Hawaii under Cooperative Agreement 
          no. NCC 5-538 with the National Aeronautics and Space Administration, Office of 
          Space Science, Planetary Astronomy Program.}
	  \and
	  D. de Winter\inst{3}
	  \and
	  B. W. Bopp\inst{4}
          }

   \offprints{M.R. P\'erez}

   \institute{Los Alamos National Laboratory, P.O. Box 1663, ISR-DO, Mail Stop F650, 
               Los Alamos, NM 87545, U.S.A \\
              \email{mperez@lanl.gov}
         \and
             Harvard-Smithsonian Center for Astrophysics, 60 Garden Street, MS 42, Cambridge, 
                MA 02138, U.S.A. \\
             \email{mvandena@eso.org}        
        \and
             TNO-TPD, Stieltjesweg 1, P.O. Box 155, 2600 AD Delft, The Netherlands\\ 
	     \email{dolf@xiada.ft.uam.es}    
	\and
             The University of Toledo, Dept. of Physics and Astronomy, Toledo, 
             OH 43606, U.S.A.\\
	      \email{bbopp@uoft02.utoledo.edu}
             }

   \date{Received September 24, 2003; November 17, 2003 accepted}

   \abstract{The southern emission-line star HD 144432 has received considerable attention due to its relative brightness (V $\sim$ 8.17), its late spectral type (late A-type or perhaps early F) and its relative isolation from a bona-fide active star formation region. We present new imaging and spectroscopic data of this star, which in the past has been classified as both evolved (post-AGB) object and an isolated Herbig Ae/Be star.  We confirm the presence of a faint companion source located 1.4\arcsec\ north, which appears physically associated with HD 144432. New infrared spectroscopy reveals this companion to be a late-type (early-mid K) star, devoid of any emission lines. Furthermore, we confirm the pre-main sequence nature of this object, report the detection of Li\,{\sc i} 6707.8\,\AA\ absorption toward the HD 144432 system, and its apparent association with Sco OB2-2 located at 145 pc.
   }
   \maketitle

%

\section{Introduction}

The star HD 144432 (SAO 184124, He 3-1141, IRAS 16038$-$2735) is one the latest in 
spectral type among the Herbig Ae/Be (HAeBe) stars and candidates listed by 
Th\'e et al. (1994). These late-type Herbig star have recently spurred some 
interest, since there appears to be a pronounced absence of emission-line stars 
with late A or early F spectral types within the group of HAeBe stars.  Since 
Herbig (1960), in the descriptive paper of the class, restricted his analysis to 
Be- and Ae-type stars, this was thought to be primarily due to historic reasons. 
However, it has proven to be difficult to find additional young objects among the 
F-type stars. Suchkov et al. (2002) from a sample of about 900 metal rich reddened 
F stars, found that about 70 stars presented abnormal near-infrared color excesses, 
suggesting the presence of circumstellar material. However, they listed only 21 
selected stars from the original large sample; HD 144432 was included among the 
several ``re-discovered'' objects.

HD 144432 was described as an IRAS point source by Oudmaijer et al. (1992), 
with the note that this object could be either a post AGB or an 
Ae Herbig star. Based on the IRAS colors, Oudmaijer et al. (1992) were 
able to derive a color or dust temperature of 240 K, by means of a 
black body fit to the excess fluxes.

Likewise, Carballo, Wesselius \& Whittet (1992), using IRAS 
fluxes, described HD 144432 as part of the Scorpio-Centaurus-Lupus region.
However, they classified it as an evolved galactic object of spectral 
type F0IIIe, with a dust shell signature due to mass loss. Furthermore, 
these authors commented that this object was suggested to be a binary by 
Wackerling (1970) and that Walker \& Wolstencroft (1988) erroneously 
classified it as a Vega-type star. 
However, in their catalogue of HaeBe stars, Th\'e et al. (1994) 
considered HD 144432 to be a probable member of this stellar group; 
a classification followed by most subsequent authors.

Motivated by the list of Vega-type stars by Walker \& Wolstencroft 
(1988), Walker \& Butner (1995) observed HD 144432, among several other 
objects, with 10 $\mu$m spectroscopy and sub-mm observations, finding 
that the stars in their sample show strong emission features due to 
silicate dust.  Similarly, Sylvester, Barlow \& Skinner (1995) 
reported a broad emission feature at 9.7 $\mu$m and possibly at 18 
$\mu$m, indicating the presence of silicate dust. These authors also 
pointed out that the excess emissions seen at long wavelengths, in HD 144432 
and other objects, have spectral slopes shallower than those predicted for 
Rayleigh-Jeans emission, suggesting the presence of large grains (radius 
$\geq$ 0.1 mm) and large amounts of cool material.

By studying a sample of stars surrounded by dust, which included HD 144432, 
Zuckerman, Forveille \& Kastner (1995) concluded that for stars with ages 
10$^6$--10$^7$ years, the gas and dust are well mixed and that this gas 
dissipates rapidly implying that if gas-giant planets (e.g., Jupiter, 
Saturn) are common in planetary systems, they must form more quickly than 
present models indicate.  

Recently, Maheswar et al. (2002) listed HD 144432 among the objects studied in search of an inconclusive geometrical relationship between binary position angle, polarization position angle and outflow position angle. Further infrared observations including HD 144432 as one the sources studied at 10 $\mu$m by van Boekel et al. (2003), found a correlation between the strength of the silicate emission, its shape and the local continuum, which is interpreted as the evidence for removal of small grains (0.1 $\mu$m), whereas large grains (1--2 $\mu$m) remain on the disk surface. 

In this paper, we present unpublished observations of the emission line 
object HD 144432, and we analyze some possible scenarios of its current 
status and evolution.
 

\section{Observations}

\subsection{Optical and near-IR Imaging}

Deep CCD images of HD 144432 through $Bessell~R$, $Gunn~i$, and narrowband 
H$\alpha$, [O\,{\sc iii}] and [Si\,{\sc ii}] filters were obtained, under 
good seeing conditions, with the Dutch 92 cm telescope at the European Southern Observatory (ESO)- La Silla, Chile on July 29, 1994. The total field of view in these images was 4'$\times$4'. Since the main purpose of these images was to look for extended, faint nebulosity in the area surrounding HD 144432, the few 
arcseconds in the immediate vicinity of HD 144432 were saturated in all images.  
No traces of extended nebulous structure or a dark region could be seen in 
any of the images, nor were there any anomalies in the relative strength in 
the various filters of the different sources in the field of view.  From 
these images we conclude that HD~144432 is apparently an isolated star, 
not directly associated with a clearly identifiable star-forming region, 
or a group of fainter emission-line stars.

\begin{figure}
   \centering
\includegraphics[angle=-90,width=8.3cm]{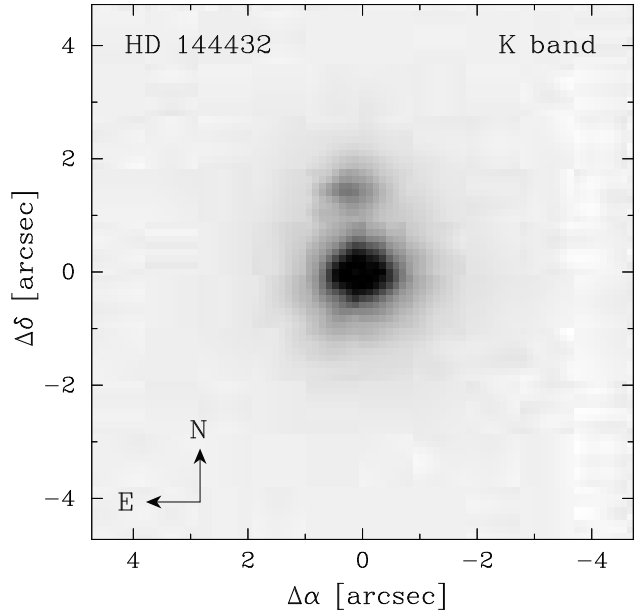}
	\caption{$K$-band (2.2 $\mu$m) image of HD~144432, showing the 	faint binary companion (HD 144432~B) located 1.4\arcsec\ mostly north from 	HD 144432~A.}
         \label{Fig1}
   \end{figure}

Recent $K$ (2.2 $\mu$m) band imaging secured at the 3.0-m NASA Infrared Telescope Facility (IRTF), reveals a faint second source (about 2.4 fainter in $K$ than the primary), located 1.4\arcsec\ north (position angle $-$7$^\circ$).  In the remainder of this paper, when distinguishing individual components is possible, we refer to the southernmost, IR-brightest component as 
HD 144432~A and the fainter, northernmost of the two, as HD 144432~B.  In 
Section 3 we argue that this secondary is physically associated with 
HD~144432~A, and we will discuss its properties.  A blow-up of the IRTF $K$-band 
image is shown in Fig.~1.  No other strong $K$-band sources were found in the 
immediate vicinity of HD~144432, confirming the earlier conclusion that the 
system appears to be rather isolated.

\subsection{Optical Spectroscopy and Li\,{\sc i} Detection}

A low-resolution (2.83~\AA~pixel$^{-1}$) CCD spectrum of HD 144432, covering 
the wavelength range of 3380 to 9180~\AA, was obtained on June 27, 1996, 
using the Boller \& Chivens spectrograph mounted on the ESO 
1.5-m telescope at La Silla, Chile. A slit 
with a width of 2\arcsec\ was used to remove the light of other parts of the 
sky. The spectrum was reduced in a standard fashion using MIDAS. Sky subtraction 
was achieved by subtracting a third degree polynomial fitted to 100 pixels 
(corresponding to roughly 25\arcsec) in the spatial direction on both sides of 
the stellar spectrum. The reduced spectrum is shown in Fig.~2.

\begin{figure}
   \centering
 \includegraphics[angle=-90,width=8.0cm]{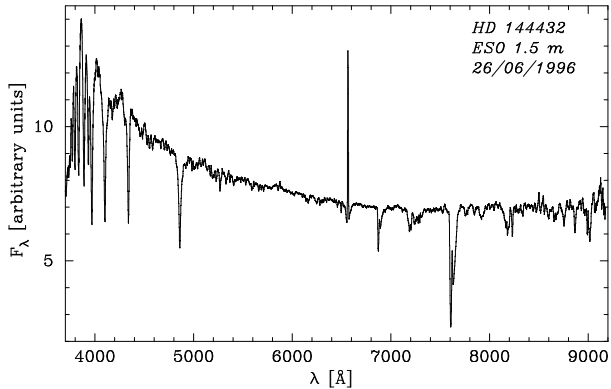}
      \caption{Optical spectra showing a strong H$\alpha$, He\,{\sc i} (5876 \AA), and Ca\,{\sc ii} triplet in emission. Continuum flux flattened out toward longer wavelengths illustrates the onset of the near-IR excess.  
              }
         \label{Fig2}
   \end{figure}

Medium resolution ($\Delta \lambda= 0.2$~\AA) spectroscopic observations were carried out at the KPNO and at Ritter Observatory, University of Toledo, at three spectral settings centered at 6430, 6750 and 8600 \AA, respectively. In the first spectral region presented in Fig.~3, metal lines in the spectrum of HD 144432 and $\sigma$ Boo (F2V) are compared. Based on the broadening of the observed lines $v \sin i$ is derived to be 72 $\pm$ 5 km~s$^{-1}$. We note that a subtle indication of asymmetry in the photospheric lines appears to be present but could not be confirmed at the resolutions presented here.

\begin{figure}
   \centering
 \includegraphics[angle=0,width=7.6cm]{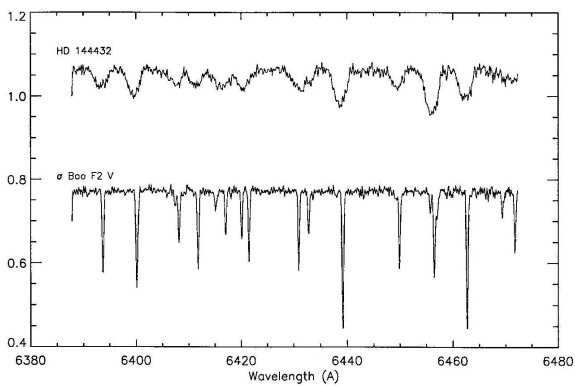}
      \caption{Metal lines in absorption compared with the main sequence $\sigma$ Boo (F2V) star. The derived rotational velocity, $v \sin i$, is 72 $\pm$ 5 km~s$^{-1}$.
              }
         \label{Fig3}
   \end{figure}

\begin{figure}
   \centering
 \includegraphics[angle=0,width=7.4cm]{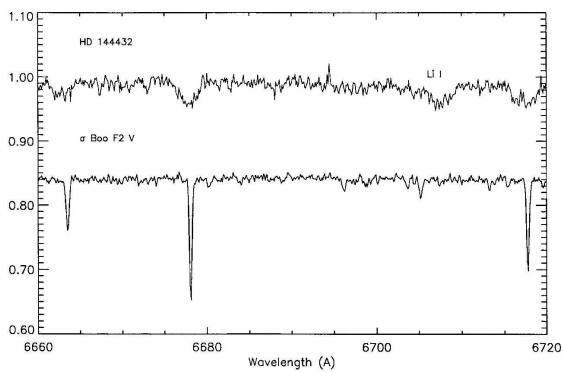}
      \caption{Lithium absorption present in HD 144432, along with the comparison star $\sigma$ Boo (F2V). Also note that Ca\,{\sc i} 6717 \AA\ also appears in absorption.
              }
         \label{Fig4}
   \end{figure}

In the second spectral region it is possible to unequivocally detect, for the first time, the absorption at Li\,{\sc i} 6707.8\,\AA\ in the spectrum of HD 144432. In Fig.~4, we present the spectral comparison of the narrow lines $\sigma$ Boo and HD 144432. We also note that Ca\,{\sc i} 6717\,\AA\ appears in absorption. Both absorption lines could suggest the presence of a cooler companion (Corporon \& Lagrange, 1999), but this suggestion is problematic as it is discussed in the conclusions section. The equivalent width of the Li\,{\sc i} line is 93 m\AA, while the nearby Fe I line 6677\,\AA\ is also 93 m\AA, both lines are visible in Fig.~4. This equivalent width of the Li\,{\sc i} line is typical of the solar-type stars in the Pleiades which are much older objects ($\sim$ 10$^{7}$ years) with significantly lower effective temperatures (e.g., Soderblom et al. 1993).

\begin{figure}
   \centering
 \includegraphics[angle=0,width=7.5cm]{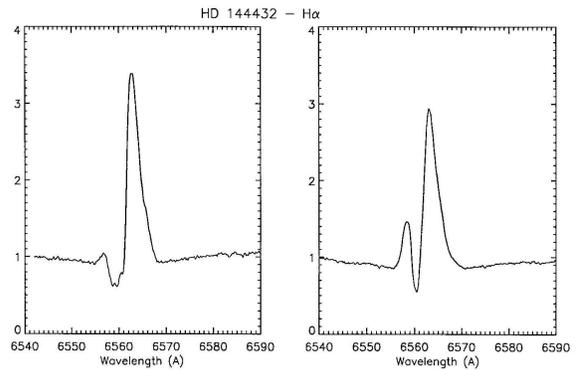}
      \caption{P Cygni H$\alpha$ emission presents season-to-season variability common in PMS objects. In contrast, photometric variability is small ($\Delta V < 0.05$ mag).}
         \label{Fig5}
   \end{figure}

Optical observations were also made of the H$\alpha$ emission. Fig.~5 presents the P Cygni profiles of HD 144432 taken in two different seasons, showing a variable blue emission component with a variable terminal velocity of $\sim$ 300-400 km/s. This strong blueshifted emission, present at both epochs, strengthens the PMS origin of this star by ruling out the possible post-AGB phase since fast winds of this kind are not observed in these stars. 

In the region centered at 8600~\AA, one can detect the presence of the Ca\,{\sc ii} infrared triplet (IRT) in emission. There are also weak Paschen absorption lines present. In Fig.~6 we have plotted the spectrum of the B8V star $\beta$ Lib to show where the Paschen wavelengths are; we do not suggest that the spectrum of HD 144432 resembles a B8V. The Ca\,{\sc ii} IRT in emission is a known signature of a shell (e.g., Jaschek et al. 1991), often present in Be stars, but beyond A0 no Ca\,{\sc ii} emission is seen, except for the alleged peculiar and young stars HD 31648 and 17 Lep (Jaschek et al. 1988), but this is a well-known characteristic of HAeBe stars, even for objects later than A0 (e.g., Catala et al. 1986, Hamann \& Persson 1992). Therefore, the presence of the Ca\,{\sc ii} IRT in emission in a A9/F0 star is peculiar but not unusual in PMS objects. 

\begin{figure}
   \centering
 \includegraphics[angle=0,width=8.0cm]{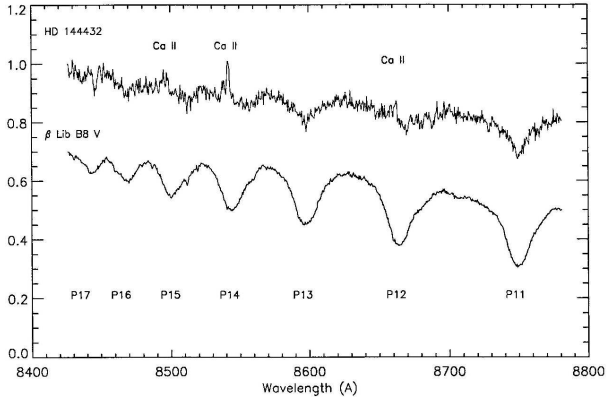}
      \caption{Ca\,{\sc ii} triplet emissions are detected in HD 144432. These lines are the signature of a shell, frequent in Be stars, but rarely present in stars later than A0. Comparison star $\beta$ Lib presents none of these emission, however, illustrates the presence of the Paschen absorption lines from P11 to P17.}
         \label{Fig6}
   \end{figure}

\subsection{Ultraviolet Spectroscopy}

HD~144432 was observed with the {\it International Ultraviolet 
Explorer (IUE)} satellite on several occasions in 1990 and 1995. Since  
for the first IUE observations, few photometric and spectroscopic observations were 
available for this object, no precise estimation of the extinction and consequently 
of the exposure times could be made in advance. Optimum 
exposure times were obtained empirically after trial observations in
both IUE cameras. All the data were obtained in the 
low dispersion mode ($\Delta \lambda \sim$ 6 \AA) 
in the long-wavelength (LWP) and short-wavelength (SWP) cameras through the 
large aperture (10\arcsec\ $\times$ 20\arcsec). We note that due to the size of the IUE entrance aperture, for all observations regardless of the aperture orientation, both stars (A and B) are expected to fall within this aperture, but due to brightness differences of the binary pair and the somewhat short exposures taken, is possibly the largest contributor to the observed fluxes, although we cannot completely exclude the potential contribution of   HD 144432~B, especially shortward of 1700 \AA.  Observations secured in 1995, have also been retrieved from the archives. The LWP image 30393 is the only high-dispersion image taken on this object. A journal of the observations is provided in Table 1. 

\begin{table}
\caption{\bf IUE Observations of HD 144432}
\begin{tabular}{ccccc}  \hline
{\bf Image}&{\bf Date}&{\bf Exp.}&{\bf V} &{\bf 
Comment} \\ 
{\bf Number}&&{\bf Time}&{\bf (FES)}&  \\ \hline
{\bf SWP} &&&& \\
38425&23 Mar 90&30 min&8.28&$\sim$optimum \\
39496&20 Aug 90&30 min&8.32&$\sim$optimum \\
54311&06 Apr 95&15 min& ---& underexposed\\
~&~&~&~&~  \\ \hline
{\bf LWP} &&&& \\
17585&23 Mar 90&10 min&8.29&over. ($\sim$ 3X) \\
17586&23 Mar 90&3 min&8.31&$\sim$optimum \\ 
30188&09 Mar 95&35 min& --- & overexposed\\ 
30354&31 Mar 95&10 min& --- & overexposed\\
30393&06 Apr 95&63 min& --- &high-disp\\
30406&08 Apr 95&1.5 min& --- & underexposed \\ \hline
\end{tabular}
\end{table}

\subsection{Near-Infrared Spectroscopy}
\begin{figure}
   \centering
 \includegraphics[angle=0,width=8.3cm]{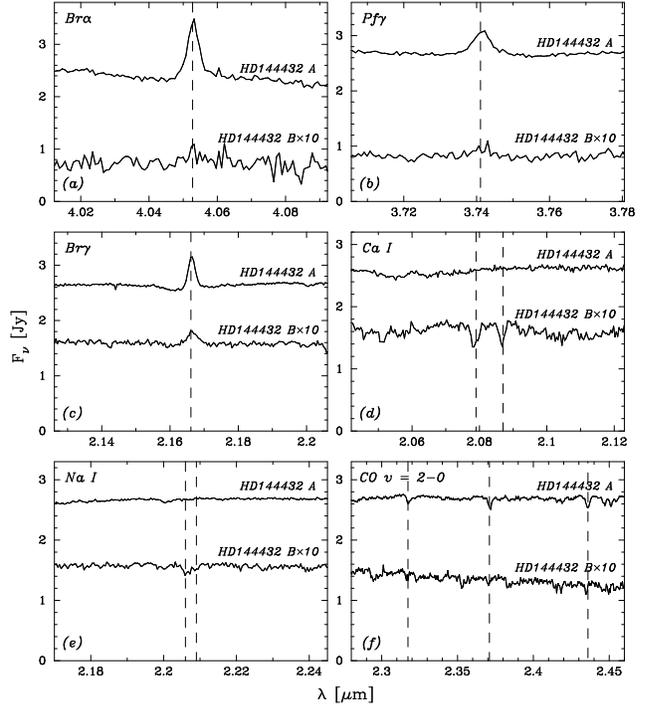}
      \caption{Overview of near-infrared emission and absorption lines detected 
in HD 144432~A and B.  (a) the Brackett $\alpha$ line at 4.05~$\mu$m, (b) Pfund $\gamma$ 
(3.74~$\mu$m), (c) Brackett $\gamma$ (2.17~$\mu$m), (d) the Ca\,{\sc i} lines at 2.08 and 
2.09~$\mu$m, (e) The Na\,{\sc i} doublet at 2.21~$\mu$m, (f) the $^{12}$CO complex 
at 2.29-2.46~$\mu$m.  The top curve in each panel shows the HD 144432~A spectrum, 
whereas the bottom curve depicts the HD 144432~B spectrum, with the flux multiplied 
by 10 for visualisation.  Dashed lines in each panel show the rest wavelength(s) of 
the depicted line(s).  Note that for the CO lines, these dashed lines indicate the 
wavelength of the band-head; absorption in the $^{12}$CO R-branch shortward of 
the band-head wavelengths is present in HD 144432~B.}
         \label{Fig7}
   \end{figure}
Near-infrared (1.9--4.1 $\mu$m) spectroscopy of the HD~144432 
system was obtained on June 28, 2002, at Mauna Kea 
Observatory with the SpeX spectrograph (Rayner et al. 2003) 
attached to the IRTF.  
Since two sources were visible on the acquisition image, 
separate spectra were taken, of the southernmost, IR-brightest, 
component, HD 144432~A, and the fainter, northernmost, of the two, 
HD 144432~B.  During both observations, the slit, with a width 
of 0.8$\arcsec$, was aligned East-West in order to minimize 
contamination of the individual spectra by the companion.

\begin{figure}
   \centering
   \includegraphics[angle=-90,width=8.3cm]{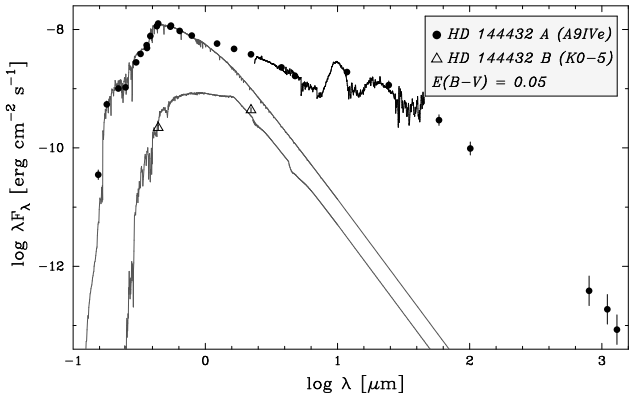}
     \caption{Spectral energy distributions (SEDs) of HD 144432~A (filled dots) and HD 144432~B (open triangles).  We also plot the ISO-SWS spectrum of HD 144432 described by Meeus et al. (2001).  For both stars we also show a reddened Kurucz (1991) model stellar photosphere fitted to the SED.  Note the strong infrared-mm wave excess above the photospheric fit for HD 144432~A.  Also note the strong emission band around 10 $\mu$m proving the presence of silicate dust. 
              }
         \label{Fig8}
   \end{figure}

The near-infrared spectrum of HD~144432~A is dominated by strong 
emission lines of H\,{\sc i}.  Br$\alpha$ (4.05 $\mu$m),
Pf$\gamma$ (3.74 $\mu$m), Br$\gamma$ (2.17 $\mu$m), and  
Br$\delta$ (1.94 $\mu$m) are clearly detected in Fig. 7.  Br$\beta$ 
(2.63~$\mu$m) is in the observed wavelength-range, but it 
is obscured by terrestrial atmospheric bands. 
The only clearly detected absorption lines in HD 144432~A 
are the $\Delta v$ = 2 band-heads of $^{12}$CO at 2.32, 
2.37 and 2.44~$\mu$m.  The solid-state band at 3.3~$\mu$m 
due to Polycyclic Aromatic Hydrocarbons (PAHS), 
often detected in HAeBe stars, is not present in 
the spectrum of HD 144432~A.  We derive an upper limit of 
4.0 $\times$ 10$^{-15}$~W~m$^{-2}$ for the flux of this 
feature in HD~144432~A.  We conclude that the near-infrared 
spectrum of HD~144432~A is clearly non-stellar in origin.  
We are probably detecting here the onset of the strong infrared 
excess due to circumstellar dust, prominently visible 
in the spectral energy distribution (SED) displayed in Fig. 8, with some emission and weaker absorption features superimposed due to a gaseous inner accretion disk.

In contrast to HD~144432~A, the near-infrared spectrum of 
HD~144432~B is dominated by absorption lines that are 
probably photospheric in origin (Fig. 7).  The Na\,{\sc i} doublet 
at 2.21~$\mu$m, as well as the Ca\,{\sc i} absorption 
lines at 1.978 and 1.987~$\mu$m, are clearly detected.  
The $^{12}$CO R-branch transition of CO is weakly present 
in absorption.  The only detected emission line is Br$\gamma$ 
at 2.17~$\mu$m.  The seeing at the time of observation, 
as well as the employed background-subtraction scheme is 
such that we can exclude that this is due to contamination 
of the point-source HD 144432~A seen in the $K$-band image.

\section{Data Analysis}  

\subsection{Variability} 
The comparison of the SWP images 38425 and 39496 taken
5 months apart revealed virtually no variability in the continuum fluxes 
beyond the repeatability errors, therefore, for the rest of the analysis
these images were coadded. The same is true for the LWP images 17585 and 17586, 
which were coadded and weighted according to their exposures times; the 
overexposed part of LWP 17585 was properly flagged and not considered in 
the final spectra. We warn that if the star is variable, the fact that no
variability is detected in these images could be only a fortunate 
coincidence. 

\subsection{FES Magnitudes} By using the IUE Fine Error Sensor (FES) counts 
recorded on the IUE observing scripts and the calibration by 
P\'erez \& Loomis (1991), we were able to estimate $V$ magnitudes 
for the times at which the 1990 IUE exposures were taken at the NASA-GSFC Observatory.  Since the 1995 exposures taken at the ESA-VILSPA Observatory, FES counts were affected by the abnormal increase in scattered light seen by the telescope optics (known as the FES anomaly detected on January 22, 1991) consequently no $V$ magnitudes were derived for these exposures.  The $V$(FES) magnitudes are listed in Table~1.  The differences in $V$(FES) values are within the errors of this calibration ($\pm$ 0.03 mag) and no significance is attributed to the small discrepancies in the $V$ magnitudes. 
We note that the non-variability of HD~144432 is in agreement with earlier photometric monitoring of HD~144432 by Meeus et al. (1998) and van den Ancker et al. (1998), but confirms this behaviour on a longer time-scale than reported in those papers.

\subsection{Spectral Type} 
When analyzing our visual spectrum of HD 144432 (see Fig.~2), it can 
be recognized as that of a star of late A or early F-type with emission lines. 
The spectrum is dominated by the Balmer lines. H$\alpha$ shows strong emission, 
whereas all other Balmer lines, visible up to H12 are in absorption. The 
Ca\,{\sc ii} K line is moderately strong in absorption (the other line of 
this doublet, Ca\,{\sc ii} H, is blended with H$\varepsilon$), suggesting a 
spectral type of late A. However, the G band at 4300~\AA\ is also definitely 
present, although not very strong, suggesting a slightly later spectral type 
of F0--2. The Na\,{\sc i} doublet at 5885~\AA\ has strong emission components, 
suggesting that these lines may be similar in nature to those of HAeBe 
stars like UX Ori (e.g. Grinin et al. 1994; Grady et al. 1996). In addition to 
this, the area between 4150 and 6500~\AA\ shows many metallic absorption lines 
of Fe\,{\sc i}, Fe\,{\sc ii}, Mg\,{\sc i} and Mg\,{\sc ii}. O\,{\sc i} may be 
present in absorption at 6161, 7777 and 8440~\AA, but these could possibly also 
be attributed to other ions. At the red end of the spectrum the Paschen lines can be seen in absorption up to P20. Superimposed on P15 and P16 we
can see two lines of the Ca triplet at 8500 and 8542~\AA\ in emission. The 
third line of this triplet, at 8665~\AA, may be present in emission, but because 
it is blended with several lines of the Paschen series this is harder to 
distinguish.

The photometric catalog by Hauck \& Mermilliod (1990) gives a value of 
H$\beta$ = 2.730 for HD~144432, which corresponds to a star of 
spectral type
F0--2 III--V according to the Str\"omgren calibration by Lester, Gray \& Kurucz (1986). However, by using the Kurucz (1991) models scaled at V=8.30, 
we show in Fig. 8 that the best fit for the unredenned optical data is for the 
model with $T_{\rm eff}$ = 7,500 K, $\log g$ = 4.0 and [Fe/H] = 0.0 (thick line), by assuming a normal extinction law (i.e., R$_{V}$=3.1).
According to the Schmidt-Kaler (1982) calibration of effective temperatures,
this corresponds to a spectral type of A8 III--V.  
The main difference between the observed data and the Kurucz model is the 
absence of Mg\,{\sc ii} absorption in HD 144432, which is discussed later. The 
predicted continuum flux as derived from the Kurucz model shortward of 1700~\AA, 
is virtually zero, whereas the stellar flux is significantly higher 
showing a moderate UV flux continuum and line excesses. In comparing with
UV standard stars from Heck et al. (1984), the spectral types A8--9 are
conspicuously absent, however, compared with HD 87696 (sp type: A7V), 
HD 144432 is probably a half or a full spectral type later. 

A respectable previous estimate of the spectral type as A9/F0V is 
presented by Houk (1982) in the Michigan catalog of Two Dimensional 
Spectral Types for the HD Stars, Vol. 3. We note that no emissions or 
peculiarities are seen in the wavelength window of 3920--4900\,\AA\ 
used for this classification.  More recent spectral classifications 
of HD~144432 as A9/F0Ve (Dunkin et al. 1997), and A9IVe (Mer\'{\i}n 
\& Montesinos 2000) confirm this spectral classification suggesting that the spectral type of HD~144432 appears constant with time.

\subsection{The Emission Spectra} A detailed analysis of the line spectra 
of the coadded data for the SWP and LWP revealed the presence of interesting 
emissions.  

\subsubsection{Short Wavelengths [1200--2000 \AA]} 

\begin{figure}
   \centering
 \includegraphics[angle=0,width=8.0cm]{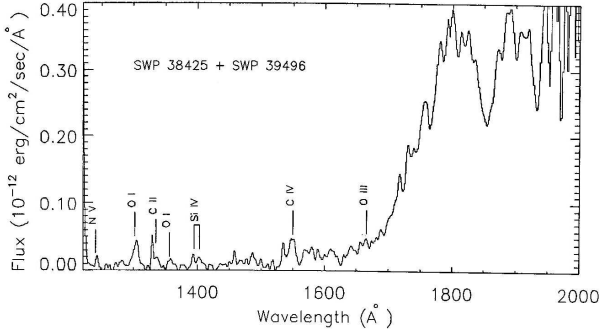}
      \caption{Short wavelength (SWP) camera spectrum showing a large number of emissions not seen in main sequence stars; N\,{\sc v}, O\,{\sc i},  O\,{\sc iii}, Si\,{\sc iv}, C\,{\sc ii}, C\,{\sc iv}, suggesting a superionization region of at least 8 $\times$ 10$^{5}$~K.
              }
         \label{Fig9}
   \end{figure}

In Fig.~9 we identified the 
emission lines present such as; N\,{\sc v} (1240\,\AA), O\,{\sc i} (1302\,\AA), 
C\,{\sc ii} (1335\,\AA), O\,{\sc i} (1356\,\AA), Si\,{\sc iv} (1394, 1403\,\AA), 
C\,{\sc iv} (1550\,\AA), O\,{\sc iii} (1666\,\AA). 
Certainly emissions at these wavelengths are not seen in main sequence
stars, however, they are detectable in stars with active chromospheres and/or
pre-main sequence objects. The star under consideration could have
an incipient chromosphere, however, this kind of activity tends to peak
at later spectral types. On the other hand, pre-main sequence stars such as
HR 5999 (spectral type A5--7 III-IV), presents the same emission lines suggesting the
presence of super-ionization regions capable of reaching temperatures
of $\sim 8 \times 10^5$ K. The origins of these emissions in the past have
been interpreted in terms of chromospheric activity, even for the earlier
types (early A- and B-type), however, recent evidences are accumulating which
favor clumpy accretion, boundary layers and bipolar flows as the main 
mechanisms responsible for most of the excesses and emission lines in the 
optical and UV (Graham 1992, Welty et al. 1992, P\'erez et al. 1993, 
Grady et al. 1993).

\subsubsection{Long Wavelengths [2000--3200 \AA]} 
In this wavelength range we can distinguish 
several Fe\,{\sc ii} UV multiplets (mostly in absorption), which are typical in
young and pre-main sequence objects. The circumstellar emission surrounding the central source is normally optically thick at UV wavelengths presenting itself as continuum flux heavily absorbed by strong Fe opacities. In 
 HD 144432, the Fe\,{\sc ii} UV multiplets 1, 62, 63 and 64 appear in absorption, however, in a few HAeBes some Fe\,{\sc ii} lines also appear in emission.  In addition, we can distinguish the emission at Mg\,{\sc ii} (UV1) and the absorption at Mg\,{\sc i} (UV1). In a study of the behavior of the Mg\,{\sc ii} lines in HAeBes, P\'erez et al. (1992) determined that the youngest and most extreme objects in the sample always presented Mg\,{\sc ii} lines (h and k) in emission, usually displaying direct P Cygni profiles, which often can only be resolved in high-dispersion ($R \sim 10^4$) images. The excess flux detection in HD 144432 of the Mg\,{\sc ii}~(2798~\AA) line from LWP images has also been reported by Valenti, Fallon \& Johns-Krull (2003).   

We also note that the resonance Mg\,{\sc ii} lines in emission are not 
unique to HAeBes and that a few stars such as B[e] stars, binaries, symbiotic 
stars and chromospherically active late-type stars also present this emission. 
The weak emission at Mg\,{\sc ii} as detected in 
low-dispersion, is a further corroboration that this object is likely to 
be young (note that classical Be and main sequence B stars generally  
present only Mg\,{\sc ii} - h and k- in absorption). Finally, the LWP high 
dispersion image, seen in Fig.~10, allows us to verify that the Mg\,{\sc ii} 
lines are indeed in emission with a well-defined P Cygni profile.

\begin{figure}
   \centering
 \includegraphics[angle=0,width=8.0cm]{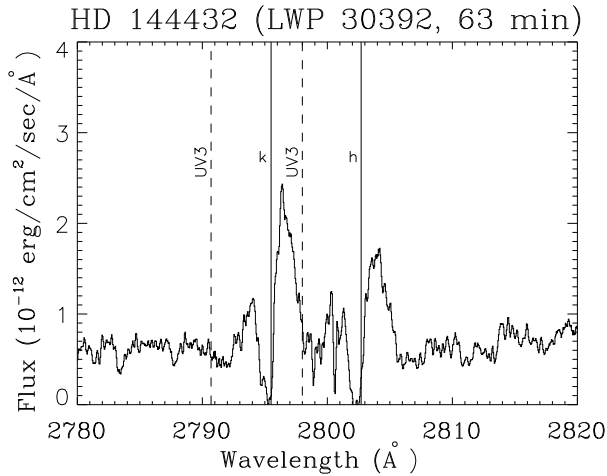}
      \caption{The P Cygni profiles of the Mg\,{\sc ii} h and k lines. The Fe\,{\sc ii} lines (UV3)  
              also are indicated.}
         \label{Fig10}
   \end{figure}

\subsection{Distance} The distance towards HD 144432 is somewhat problematic.  The distances 
quoted in the literature range from 108 pc to 2.4 kpc. The near distance determination is based 
on a so-called ``spectroscopic parallax'', i.e. simply by assuming that the star is on the 
main-sequence, which is probably incorrect.  From similar arguments, but assuming that 
HD 144432 is a post-AGB star, Pottasch \& Parthasarathy (1988) conclude that the distance 
towards HD 144432 is 2.4 kpc. 

HD 144432 is located in the region of the sky occupied by Sco OB2-2 ($d$ = 145 pc), which would 
strongly indicate that it belongs to this group, suggesting a distance of about 145 pc. Two 
B2V stars, 13 Sco and $\rho$ Sco are a few degrees away from HD 144432 and both are members of the Sco OB2-2 association. The proper motions of all three stars agree suggesting that they all belong to Sco OB2-2. 

HD 144432 was also measured by the Hipparcos satellite, but the parallax 
measurement only resulted in a 2.5 $\sigma$ detection.  However, its distance 
was derived to be 250 pc (with a large error bar); inconsistent with the 
determination by Pottasch \& Parthasarathy (1988).  
Additional evidence for a near rather than a far distance are the weak interstellar 
Na\,{\sc i} D-line (5890 \AA) shown by Meeus et al. (1998) and also observed by 
one of us (BWB).  We conclude that the observed Na\,{\sc i} D-line strength 
constrains the distance towards HD~144432 to be between 100 and 500 pc, ruling 
out the 2.4 kpc distance derived by Pottasch \& Parthasarathy (1988).

Perhaps the best estimate for the true distance may come from an astrophysical argument. 
From the spectral energy distribution, one can compute a luminosity density, $L_{\star}/d^2$, of 
4.9 $\times$ 10$^{-4}$ L$_{\sun}$~pc$^{-2}$ for HD 144432.  The luminosity for a A9 V star is 
7.6 L$_{\sun}$, which would translate to a distance of 124 pc.  If we assume HD 144432 to be an  
A9 III giant, its luminosity should be around 23 L$_{\sun}$, which would place it around 216 pc.  
So, if we assume that the spectroscopic classification of HD 144432 as A9 IV is correct, this would lead us to a distance smaller than 200 pc, which now contradicts the Hipparcos data. We note that the renewed realization that HD 144432 is a binary system also casts some doubt on the Hipparcos result, since an unrecognized binary motion in the system could affect the outcome of the solution for the parallax by the automated Hipparcos algorithm.

We conclude that HD 144432 is most likely a member of Sco OB2-2, with a 
distance of 145 pc. However, we warn about the discrepancy with Hipparcos parallax measurements, which for many other young objects have proven to be a powerful tool to ascertain their distances (van den Ancker et al. 1997, 1998).

\subsection{The Companion}
A faint companion to HD 144432 was detected in the $K$-band (2.2~$\mu$m) 
image of HD 144432, obtained at the IRTF (Fig.~1).  As was already noted 
in Section 2.4, the infrared spectrum of the companion, HD 144432~B, is 
dominated by absorption lines that are probably photospheric in origin.  
The ratio of the strength of the 2.21~$\mu$m Na\,{\sc i} doublet to that 
of the 2.30~$\mu$m CO feature has been shown to be a good indicator of 
spectral type in pre-main sequence stars (e.g. Doppmann \& Jaffe 2003).  
The fact that they appear to have nearly equal strength in HD 144432~B, 
suggests a spectral type of early-mid K for this star.

Our IRTF image shows that in June 2002, HD 144432~B was located 
1.4$\arcsec$ from the primary at a position angle of $-$7$^{\circ}$.  
A search of the SIMBAD database revealed that HD 144432 had been 
classified before as a close visual binary, Rossiter 1877 (Rossiter 1955).  
The 1934 measurement of this gave a separation of 1.2" with position 
angle of 354 degrees, and photographic magnitudes of 8.4 and 12.9 for 
the two components.  This is essentially identical to the position of 
the companion seen at the IRTF.  
With the early-K spectral type derived from the IRTF spectrum, 
the photographic magnitude of 12.9 is also compatible with the brightness 
of HD 144432~B in the IRTF $K$-band image.  Therefore, we conclude we 
are seeing the same companion, at nearly the same relative position, 
as was seen in 1934 in the optical.  

We note that the proper motion of HD 144432 as measured 
by Hipparcos (30 milliarcseconds per year), means that in the course of 
the 68 years between the two measurements, the separation between the two 
should have changed by about 2.0\arcsec\ if the two components had different 
proper motions.  This has apparently not happened.  Consequently, we conclude 
that HD 144432~B shares a common proper motion with HD 144432~A, strongly suggesting a common distance and origin.

From the detailed SED of HD 144432~A (Fig.~8), we derive $L_{\star}/d^2$ 
of 7.1 $\times$ 10$^{-4}$ L$_\odot$~pc$^{-2}$ 
for HD 144432 A.  The IRTF $K$ band magnitude of 8.3 for HD 144432~B 
translates to $L_{\star}/d^2$ = 4.4 $\times$ 10$^{-5}$ L$_\odot$~pc$^{-2}$ 
if we adopt a K2 spectral type and assume the (small) 
colour excess $E(B-V)$ for HD 144432 A and B to be the same.  If we 
plot these values in a HR diagram for the two possible distances of 145 
and 250 pc (Fig. 11), we note that for both distances, HD 144432~A and B would be located above the main sequence, in an area typically occupied by pre-main sequence stars.  Although we cannot exclude 
the 250~pc distance completely, we note that for the 145 pc 
distance both HD 144432 A and B are located between the isochrones 
for 1 and 3 million years, whereas for the 250 pc, a small discrepancy 
in age between the two systems would exist.

\section{Conclusions}
Our observations have demonstrated that HD 144432 is indeed a complex and intriguing object. We confirm its pre-main sequence nature, and have verified the existence of a faint late-type companion to HD 144432.  The common proper motion and age of 
HD 144432 A and B strongly suggest that they have formed together.  Further 
research is needed to investigate whether this companion, as HD 144432~A itself, 
is still surrounded by significant amounts of circumstellar material, or 
whether it has already shed most of its natal cloud.

The detection of Li\,{\sc i} (6708 \AA) absorption in the integrated spectrum 
of HD 144432 is intriguing.  The simple-minded explanation that this could form 
in the atmosphere of the late-type star HD 144432~B is problematic, since 
the Li\,{\sc i} line presents the same rotational broadening (see Fig.~4) as the 
metal lines.  Unless the low-mass binary companion has a similar rotational 
broadening as the primary, which we consider unlikely, this strongly suggests 
that Li\,{\sc i} is produced in the same stellar photosphere, i.e., intrinsic 
to HD 144432~A.  Early F-type stars can present the Li\,{\sc i} absorption, 
but at this photospheric temperature Li is mostly in the form of Li\,{\sc ii}, 
so our measured EW(Li\,{\sc i}) of 93 m\AA\ would imply an usual high Li 
abundance in HD 144432. 

The UV observations of HD 144432 also confirmed the youth of this star, as a 
probable pre-main sequence object. The UV spectral type of this
star appears to be earlier than the optical type. The trend of earlier 
spectral types with decreasing wavelength is a clear indication of the 
inconsistency of the single-temperature stellar model, which has been found 
to be the hallmark of young objects such as FU Orionis (Hartmann \& Kenyon 
1985) and other HAeBe stars (P\'erez et al. 1993, and references 
therein). This effect has been interpreted as evidence for the presence of 
emission from an optically thick accretion disk and boundary layer, rather
than from a stellar photosphere.  The 
presence of several emission lines shortward of 1700 \AA, species commonly 
associated with chromospheres in late-type stars, suggests that these 
emissions probably have a more energetic origin such as the one predicted by 
the accretion disk + boundary layer model for HAeBe stars (Hillenbrand 
et al. 1992; Blondel et al. 1993). Corroborating evidences for the youth of
this object are its physical location in the upper Scorpius region and the
large infrared excess indicating the existence of large amount of circumstellar
material capable of thermally re-emitting at longer wavelengths. 

\begin{figure}
   \centering
 \includegraphics[angle=-90,width=8.3cm]{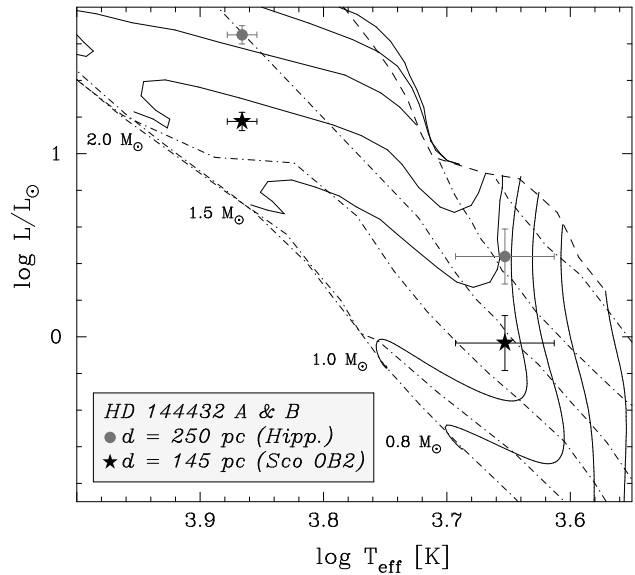}
      \caption{Hertzsprung-Russell diagram for HD 144432~A and B for a common distance of 145~pc (dots) and 250~pc (stars).  Also shown are the pre-main sequence evolutionary tracks (solid lines) and isochrones (dash-dotted) by Palla \& Stahler (1993).}
         \label{Fig11}
   \end{figure}

\begin{acknowledgements}
      This research made use of the SIMBAD database, operated at CDS, Strasbourg, France. We are grateful to Dr. Fancis Fekel for obtaining spectra at 6430 \AA\ (Fig. 3) at our request.  We are especially thankful to the referee of this manuscript, Dr. Claude Catala, for the several suggestions that resulted in many clarifications and improvements to this paper. We also acknowledge the financial assistance of the Los Alamos Laboratory-Directed Research and Development (LDRD) used to complete work included in this paper (LA-UR-03-6569). 

\end{acknowledgements}

\end{document}